\documentclass[final,3p,times]{elsarticle}

\usepackage{wrapfig}

\usepackage{graphics}
\usepackage{graphicx}
\usepackage{epsfig}
\usepackage{amssymb}
\usepackage{amsthm}
\usepackage{amsmath}

\usepackage{caption}
\usepackage[labelfont=bf]{caption}
\usepackage[labelsep=space]{caption}
\usepackage[nodots, compress]{numcompress}
\biboptions{sort&compress}

\usepackage{xcolor}
\usepackage{hyperref}
\hypersetup{
colorlinks,
citecolor=[violet],
linkcolor=[red],
urlcolor=[blue]
}

\journal{Materials Today Communications}

\begin{document}

\begin{frontmatter}

\title{A new ferromagnetic semiconductor system of Eu$_{1-x}$Sr$_x$AgP $(x = 0.0-0.6)$ compounds: Crystallographic, magnetic, and magneto-resistive properties}

\author[1]{Qian Zhao}
\author[1]{Kaitong Sun}
\author[2]{Junchao Xia}
\author[1]{Hai-Feng Li\corref{cor1}}
\cortext[cor1]{Corresponding author}
\ead{haifengli@um.edu.mo}

\affiliation[1]{organization={Institute of Applied Physics and Materials Engineering, University of Macau},
            addressline={Avenida da Universidade, Taipa},
            city={Macao SAR},
            postcode={999078},
            country={China}}
\affiliation[2]{organization={School of Materials Science and Engineering, Chang’an University},
            city={Xi’an },
            postcode={710061},
            country={China}}

\begin{abstract}
Adjusting chemical pressure through doping is a highly effective method for customizing the chemical and physical properties of materials, along with their respective phase diagrams, thereby uncovering novel quantum phenomena. Here, we successfully synthesized Sr-doped Eu$_{1-x}$Sr$_x$AgP $(x = 0.0-0.6)$ and conducted a comprehensive investigation involving crystallography, magnetization, heat capacity, and magnetoresistance. Utilizing X-ray diffraction and PPMS DynaCool measurements, we studied Eu$_{1-x}$Sr$_x$AgP in detail. The hexagonal structure of parent EuAgP at room temperature, with the $P6_3/mmc$ space group, remains unaltered, while the lattice constants expand. The magnetic phase transition from paramagnetism to ferromagnetism, as temperature decreases, is suppressed through the gradual introduction of strontium doping. Heat capacity measurements reveal a shift from magnon-dominated to predominantly phonon and electron contributions near the ferromagnetic phase with increasing doping levels. The resistivity-temperature relationship displays distinct characteristics, emphasizing the impact of Sr doping on modifying charge transport. Magnetoresistance measurements uncover novel phenomena, illustrating the adjustability of magnetoresistance through Sr doping. Notably, Sr doping results in both positive magnetoresistance (up to 20\%) and negative magnetoresistance (approximately -60\%). The resistivity and magnetic phase diagram were established for the first time, revealing the pronounced feasibility of Sr doping in modulating EuAgP's resistivity. This study has provided valuable insights into the intricate interplay between structural modifications and diverse physical properties. The potential for technological advancements and the exploration of novel quantum states make Sr-doped Eu$_{1-x}$Sr$_x$AgP a compelling subject for continued research in the field of applied physics.
\end{abstract}

\begin{graphicalabstract}
\includegraphics[width=0.88\textwidth]{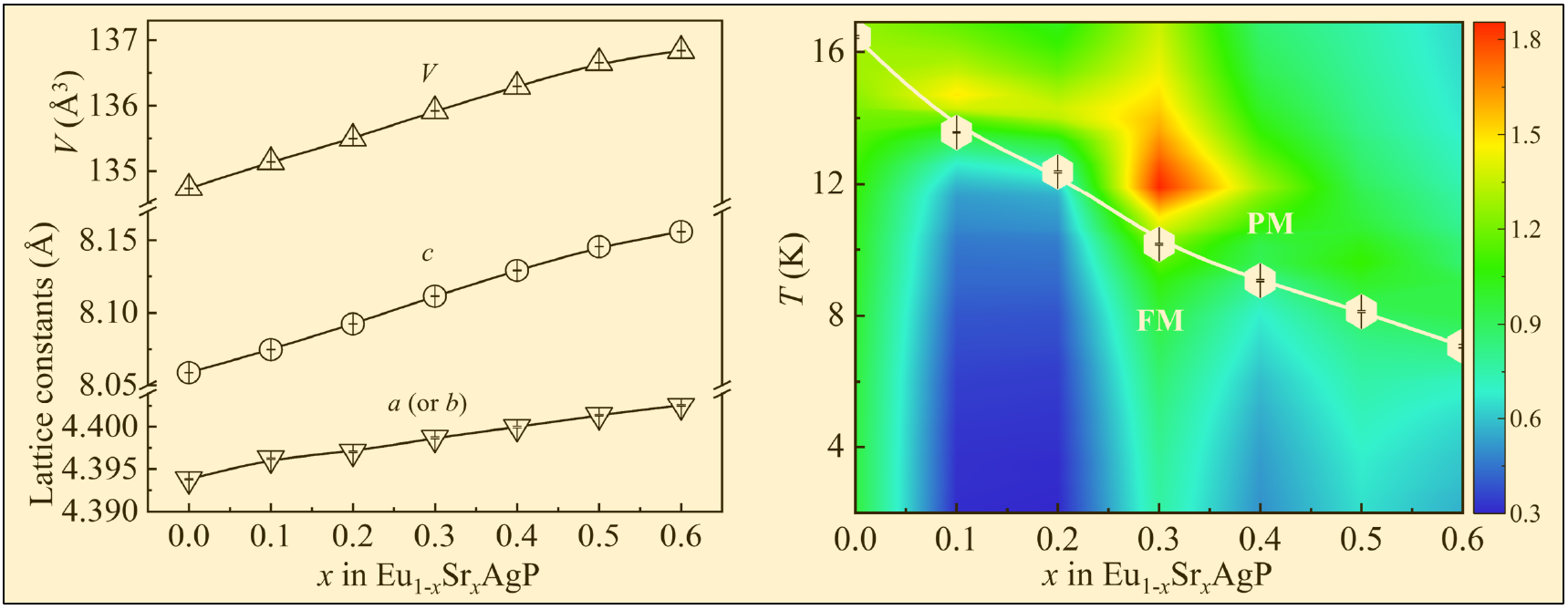} \\
\medskip
\noindent{\textbf{Caption of Graphical Abstract:} (left) Sr-Doping Tunability: The successful incorporation of Sr into EuAgP leads to the expansion of lattice parameters. (right) Comprehensive Phase Diagram: For the first time, we have established the resistivity and magnetic phase diagram for Eu$_{1-x}$Sr$_x$AgP $(x = 0.0-0.6)$ compounds.}
\bigskip
\label{GA}
\end{graphicalabstract}

\begin{highlights}
\item Successfully incorporated Sr into EuAgP, expanding lattice constants.
\item Sr doping suppresses ferromagnetic transition temperature systematically.
\item Sr doping leads to both positive (up to 20\%) and negative (-60\%) magnetoresistance.
\item First-time establishment of resistivity and magnetic phase diagram.
\item Opens avenues for innovative magnetic memory devices and sensor/actuator applications.
\end{highlights}


\end{frontmatter}


\section{Introduction}

Europium-based rare-earth compounds have long fascinated researchers due to their intricate and innovative physical properties.\cite{ZHAO2024,shen2023magnetic, gao2021time, nie2020magnetic, ma2020emergence} These materials encompass topological compounds, magnetic compounds, as well as insulators, semiconductors, semi-metals, and half-metals.\cite{takahashi2018anomalous, ma2019spin, shen2023magnetic, xu2021unconventional, tong2014magnetic} They can be finely manipulated in their spin and electronic states through various methods such as crystal structure manipulation, pressure, magnetic fields, and doping, allowing for the exploration of novel physical properties and interactions. The versatility of these materials, ranging from topological compounds with unique electronic band structures to magnetic compounds exhibiting intriguing magnetic phenomena, has ignited extensive research efforts. Recently, considerable attention has been focused on ternary compounds denoted as Eu$TX$, where $T$ includes Cu, Ag, Au, and $X$ includes P, As, Sb. Notably, within this class of compounds, Eu$T$P compounds exhibit ferromagnetic (FM) characteristics,\cite{WANG2023169620} while Eu$T$As/Sb compounds manifest antiferromagnetic behavior.\cite{ZHANG2023101228, laha2021topological, jin2021multiple} Zhang \emph{et al}. investigated the high-pressure characteristics of the antiferromagnetic Dirac semimetal EuAgAs single crystal and identified a pressure-induced structural phase transition leading to a spin configuration reconstruction, resulting in a quasi-FM state.\cite{ZHANG2023101228} Wang \emph{et al}. studied the hexagonal FM single crystal EuCuP, revealing an FM transition at $T_{\textrm{C}}$ = 32 K accompanied by the manifestation of an anomalous Hall effect.\cite{WANG2023169620} EuAgP was initially synthesized in 1981, initiating investigations into its crystal structure and magnetism.\cite{tomuschat1981abx} In recent years, first-principles calculations have predicted EuAgP to belong to the category of magnetic semiconductors or Weyl semimetals.\cite{pottgen2000equiatomic} More recent experimental studies have conducted detailed examinations, revealing the actual band gap value and showcasing rich magneto-resistive phenomena.

The incorporation of strontium (Sr) has found widespread application in the modulation of electronic states, as evidenced by early studies on LaMnO$_3$ doping\cite{phan2005excellent, phan2010tricritical, jeddi2018magnetocaloric, TABOADAMORENO2020165887} and investigations into Ni-based superconductors.\cite{li2019superconductivity, ryee2020induced, li2020absence, Osada2021Nickelate} However, in europium (Eu)-based transition metal compounds, Sr doping remains unexplored. The Sr$^{2+}$ ion exhibits a remarkably close ionic radius to Eu$^{2+}$ when the coordination number is the same.\cite{Shannon:a12967} Therefore, doping can readily occur through a substitutional mechanism. Sr$^{2+}$ is commonly regarded as a nonmagnetic ion due to its electronic configuration and magnetic properties. The Sr$^{2+}$ ion possesses a fully filled 4$s^2$ electron shell, devoid of any unpaired electrons. According to quantum mechanics, the presence of unpaired electrons in an atom or ion is essential for manifesting magnetic properties, as these unpaired electrons can align their spins in response to an external magnetic field, resulting in magnetic moments. Since Sr$^{2+}$ lacks any unpaired electrons, it does not exhibit inherent magnetic moments and is therefore classified as nonmagnetic. Consequently, when Sr is introduced into a material through doping, it typically does not directly contribute to the material's magnetic behavior. Instead, its influence often lies in altering the electronic and structural properties of the material, thereby indirectly affecting its magnetic characteristics. Under such conditions, the structural integrity of Eu$TX$ is preserved, and Sr$^{2+}$, being a nonmagnetic ion, can modulate the magnetic properties of the sample. This approach allows for the manipulation of the magnetic transition temperature, aiming to broaden their potential applications of the samples. Sr doping may also induce unique electronic states.

In this study, we successfully synthesized a series of polycrystalline samples of Eu$_{1-x}$Sr$_x$AgP with $x$ ranging from 0.0 to 0.6. The single-phase nature and effectiveness of doping were confirmed through room temperature X-ray powder diffraction (XRPD). A comprehensive investigation of magnetic, heat capacity, and electrical transport properties was carried out using a physical property measurement system (PPMS DynaCool instrument, Quantum Design) for this series of samples. The research focused on the magnetic and magneto-resistive effects under the influence of doping. Novel positive and enhanced negative magnetoresistance effects were observed for the first time, and temperature-dependent magnetic and resistivity phase diagrams were constructed and analyzed.

\section{Materials and methods}
\subsection{Materials preparation}

Polycrystalline powders of Eu$_{1-x}$Sr$_x$AgP $(x = 0.0-0.6)$ with diverse chemical stoichiometry ratios were synthesized. The synthesis involved the utilization of cut europium blocks with shiny surfaces, silver powder, cut strontium blocks with shiny surfaces, and phosphorus powder. After mixing and grinding in a mortar, the materials were placed into specially designed alumina crucibles, sealed within quartz tubes under an argon atmosphere.\cite{johrendt1997tuning} The sealed quartz tubes were gradually heated (50 K/hour) to 1223 K and held at that temperature for 24 hours. Subsequently, they were allowed to cool naturally to room temperature by shutting down the furnace. This pre-melting step aimed to facilitate the initial blending of the bulk metal raw material with the powder raw material. The obtained samples underwent a 24-hour annealing process at 1323 K. Following natural cooling, the samples were ball-milled for 15 minutes and then resealed in a quartz tube for the third sintering at 1373 K, holding for 36 hours. This process resulted in solid samples exhibiting a dark-grey color. The materials remained stable in ambient air.

\subsection{X-ray powder diffraction}

The resulting materials was finely milled into a powdered sample using a Vibratory Micro Mill (FRITSCH PULVERISETTE 0) for structural analysis. XRPD was conducted at room temperature over a 2$\theta$ range of 15--85$^\circ$, with a step size of 0.02$^{\circ}$ on our in-house X-ray diffractometer (Rigaku, SmartLab 9 kW). Copper $K_{\alpha1}$ (1.54056 {\AA}) and $K_{\alpha2}$ (1.54439 {\AA}) were employed in a 2:1 intensity ratio as the radiation source. XRPD patterns were recorded under ambient conditions at a voltage of 45 kV and a current of 200 mA. The collected XRPD patterns were refined using the FULLPROF SUITE program.\cite{fullprof}

\subsection{Magnetization measurement}

Using the DC magnetization measurement capabilities of our PPMS DynaCool instrument, we performed measurements on pressed Eu$_{1-x}$Sr$_x$AgP $(x = 0.0-0.6)$ pellets. These pellets were prepared by applying a pressure of approximately 60 MPa for 15 minutes. In the temperature range of 1.8 to 200 K, we conducted zero-field-cooling magnetization measurements on samples with varying doping concentrations under an applied magnetic field of 100 Oe. Additionally, we performed magnetic hysteresis loop measurements on samples with different doping concentrations at a low temperature of 1.8 K, ranging from 0 to 14 T, then from 14 to -14 T, and finally back to 14 T.

\subsection{Heat capacity measurement}

For the samples of heat capacity and electrical transport, the pressed pellets underwent sintering at 700 $^{\circ}$C for 24 hours prior to be measured. Utilizing the specific heat option of the PPMS, we conducted measurements on Eu$_{1-x}$Sr$_x$AgP $(x = 0.0-0.6)$ samples with varying doping concentrations at zero magnetic field, covering a temperature range from 1.8 to 200 K.

\subsection{Resistivity measurement}

The temperature-dependent resistivity of the polycrystalline Eu$_{1-x}$Sr$_x$AgP bars, cut from the corresponding pellets, was evaluated using the resistivity option of PPMS. Measurements were conducted across a temperature range of 2 to 300 K, with varying doping levels, employing a four-probe setup under magnetic fields ranging from 0 to 14 T, and SPI silver paint was utilized to link Au wires to the measured sample.

\section{Results and discussion}

\subsection{Structural study}
\begin{figure*}[t]
\centering
\includegraphics[width = 0.88\textwidth] {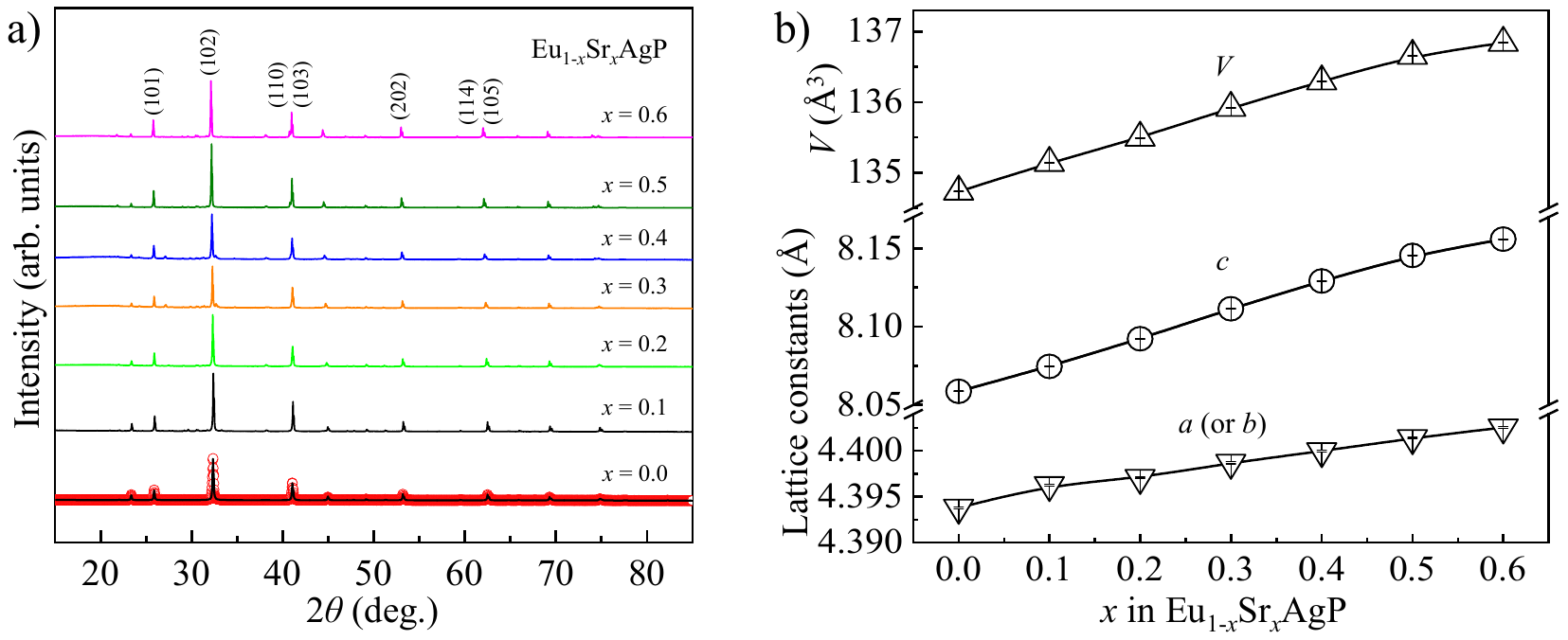}
\caption{XRPD analysis and associated refinement results.
a) XRPD patterns of Eu$_{1-x}$Sr$_x$AgP $(x = 0.0-0.6)$ at room temperature. For $x = 0$, circles represent the original data, and the solid line depicts the calculated pattern using the $P6_3/mmc$ (No. 194) space group. Various colors in the XRPD patterns indicate different doping concentrations in Eu$_{1-x}$Sr$_x$AgP samples with the doping level ranging from 0.0 to 0.6. The three numbers in parentheses denote the major Bragg peaks.
b) Lattice constants $a$ (= $b$) and $c$, along with the corresponding unit-cell volume $V$, for Eu$_{1-x}$Sr$_x$AgP samples, refined from room-temperature XRPD data, are plotted as a function of the doping level.
}
\label{XRPD}
\end{figure*}

To validate the effectiveness of Sr doping in Eu$_{1-x}$Sr$_x$AgP $(x = 0.0-0.6)$ samples, we performed room temperature XRPD on samples with varying doping gradients to study the structural changes in Eu$_{1-x}$Sr$_x$AgP after doping. Considering the coordination number of 8 for both Eu$^{2+}$ and Sr$^{2+}$ ions and their respective ionic radii of 1.25 and 1.26 Å, occupying the same crystallographic 2a position is expected. Thus, Eu$_{1-x}$Sr$_x$AgP doped with Sr maintains the same structure as the parent EuAgP ($P6_3/mmc$). As depicted in Fig.~\ref{XRPD}a, the XRPD patterns of the doped samples exhibit excellent phase purity and align with EuAgP, confirming their single-phase nature.\cite{tomuschat1981abx, johrendt1997tuning} The major Bragg peaks show no significant shifts or splitting, indicating the overall structural integrity of the doped samples. Refinement of all XRPD data using the $P6_3/mmc$ space group indicated that the structure of Eu$_{1-x}$Sr$_x$AgP $(x = 0.0-0.6)$ remains unchanged after doping, with Sr$^{2+}$ partially occupying the original Eu$^{2+}$ position. Notably, the Bragg peaks near 2$\theta$ = 75$^\circ$ exhibit slight shifts and splitting as the doping level increases. This effect may be attributed to the possible substitution of Sr$^{2+}$ into alternative sites within the EuAgP crystalline structure, causing possible structural distortions.\cite{TABOADAMORENO2020165887}

We performed careful refinements of all XRPD data using FULLPROF software (Fig. S1) and subsequently extracted lattice constants and associated parameters. The unit cell of the resulting crystal structure is depicted in Fig. S2. Due to the slightly larger size of Sr$^{2+}$ ions compared to Eu$^{2+}$, the lattice constants $a$ (= $b$) and $c$ of Eu$_{1-x}$Sr$_x$AgP $(x = 0.0-0.6)$ after doping should all increase. As depicted in Fig.~\ref{XRPD}b, the lattice constants $a$ (= $b$) and $c$ of Eu$_{1-x}$Sr$_x$AgP noticeably expand with increasing doping concentrations, particularly evident in the cell volume $V$, which substantiates the success and pronounced effectiveness of the Sr doping we performed on EuAgP. At the same time, this may suggest potential internal lattice distortions or internal tensile stresses.\cite{Yao2006Effects} The refined structural parameters are all listed in Table~\ref{ref-para}.

\subsection{Magnetization study}
\begin{figure*}[t]
\centering
\includegraphics[width = 0.88\textwidth] {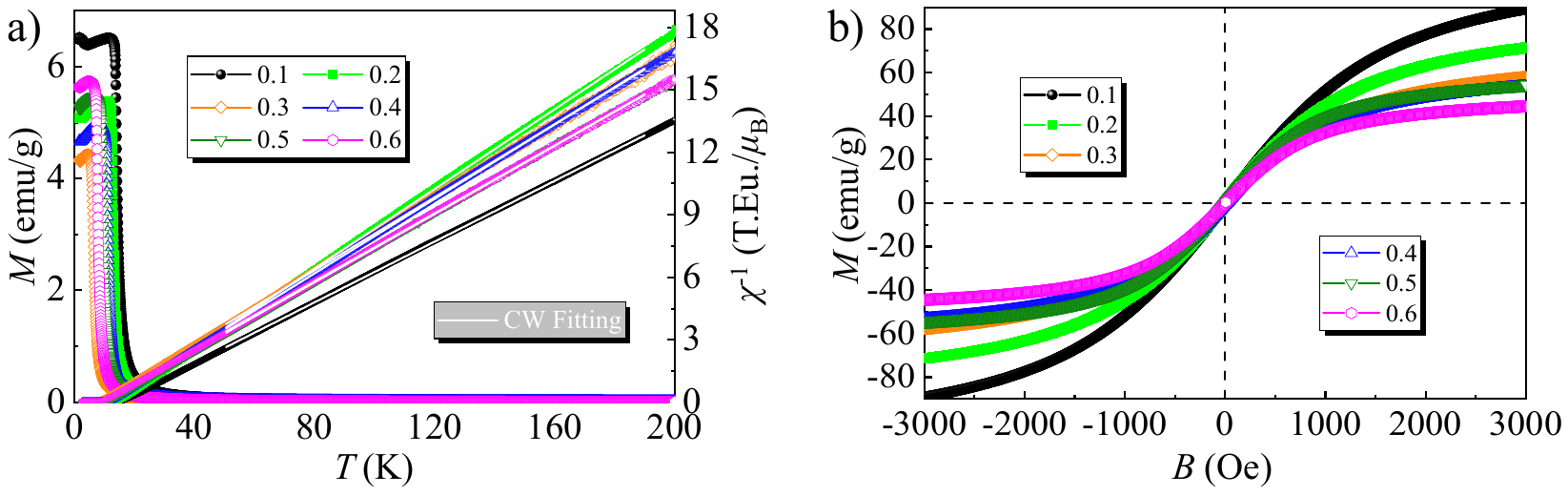}
\caption{Magnetization as a function of temperature and applied magnetic field.
a) Magnetization under the zero-field-cooling condition ($M$ curves, on the left axis) and the corresponding calculated inverse magnetic susceptibility (${ \chi }^{-1}$ straight lines, on the right axis) are presented with an applied magnetic field of 100 Oe across the temperature range of 1.8 K to 200 K. The white line represents the fit using the Curie–Weiss law.
b) The magnetization is plotted against the magnetic field ranging from -3000 to 3000 Oe, measured for various doping levels (0.1 to 0.6) of Eu$_{1-x}$Sr$_x$AgP samples at 1.8 K.
}
\label{Mag}
\end{figure*}

As depicted in Figs.~\ref{Mag}a and S3, the temperature-dependent magnetization curves cover a spectrum of Eu$_{1-x}$Sr$_x$AgP $(x = 0.0-0.6)$ samples, encompassing doping concentrations ranging from 0.0 to 0.6. Illustrated on the left axis of Fig.~\ref{Mag}a, the magnetization gradually increases as the temperature decreases from 200 to 30 K. It demonstrates a smooth ascent, commencing around 25 K and peaking at approximately 15 K, followed by a rapid decline. Concurrently, the transition temperature of magnetization exhibits noticeable variations with increasing doping levels. These curves vividly illustrate the distinct features of a magnetic phase transition associated with Eu$^{2+}$ ions. The fitting of the inverse magnetic susceptibility ${\chi}^{-1}$ in the pure paramagnetic (PM) state employs the Curie-Weiss (CW) law,
\begin{eqnarray}
\chi^{-1}(T) = \frac{3k_\textrm{B}(T - \Theta_{\textrm{CW}})}{N_\textrm{A} \mu^2_{\textrm{eff}}},
\label{CWLaw}
\end{eqnarray}
where $k_\textrm{B} =$ 1.38062 $\times$ 10$^{-23}$ J/K represents the Boltzmann constant, ${\Theta}_\textrm{CW}$ denotes the CW temperature, $N_\textrm{A} =$ 6.022 $\times$ 10$^{23}$ mol$^{-1}$ is Avogadro's constant, and $\mu{\textrm{eff}}$ is the effective PM moment defined as $g\mu_\textrm{B} \sqrt{J(J + 1)}$. The fitting process involved applying Eq.~\ref{CWLaw} to all six susceptibility curves within the temperature range of 50--190 K. Subsequently, we determined the intercept on the horizontal-axis, corresponding to $M({\Theta}_\textrm{CW}) =$ 0 (Fig.~\ref{Mag}a). It is pointed out that in the calculation of the magnetic susceptibility for Eu$_{1-x}$Sr$_x$AgP samples with different doping concentrations, we have normalized the Eu$^{2+}$ concentrations for various stoichiometries. Consequently, the idealized CW fitting curves should exhibit the same slope for different doping gradients, as per the theoretical value (7.94 ${\mu}_\textrm{B}$) of Eu$^{2+}$ ions (shell 4$f^7$, quantum numbers $S = \frac{7}{2}$, $L$ = 0, and $J = \frac{7}{2}$). However, as shown in Fig.~\ref{Mag}a, samples with $x$ = 0.1--0.6 show distinct slopes, potentially attributed to the variation in doping levels. Thus, we performed CW fittings to separately calculate the effective PM moments for different doping levels, presented in Table~\ref{ref-para}. Notably, the effective PM moment of the Eu${_\textrm{0.9}}$Sr${_\textrm{0.1}}$AgP sample is remarkably close to the ideal value. This observation suggests that the Sr doping concentration with $x =$ 0.1 is highly advantageous in reducing the occurrence of Eu$^{3+}$ impurities during the synthesis of Eu$_{0.9}$Sr$_{0.1}$AgP.

The introduction of non-magnetic Sr$^{2+}$ ions did not induce noticeable magnetic phase transitions in the Eu$_{1-x}$Sr$_x$AgP $(x = 0.0-0.6)$ samples. We still observed magnetic behaviors similar to those of EuAgP. Hysteresis loops at 1.8 K with different doping concentrations are presented in Fig.~\ref{Mag}b. As the concentration of magnetic Eu$^{2+}$ ions contributing to magnetism decreases, the saturated magnetization gradually diminishes. However, the shape of the hysteresis loop, characterized by the remanent magnetization and coercive field, remains constant with varying doping levels. Only weak loops are observable, indicating the soft FM nature, similar to that of EuAgP. At a temperature of 1.8 K and for $x$ = 0.1, where the coercive field and residual magnetization are expected to be relatively highest, the coercive field $H_\textrm{c}$ = $\sim$ 75.45 Oe, the residual magnetization $M_\textrm{r}$ = $\sim$ 0.45 emu/g, and the saturation magnetization $M_\textrm{s}$ = $\sim$ 113.32 emu/g.

\subsection{Heat capacity study}
\begin{figure}[t]
\centering
\includegraphics[width = 0.48\textwidth] {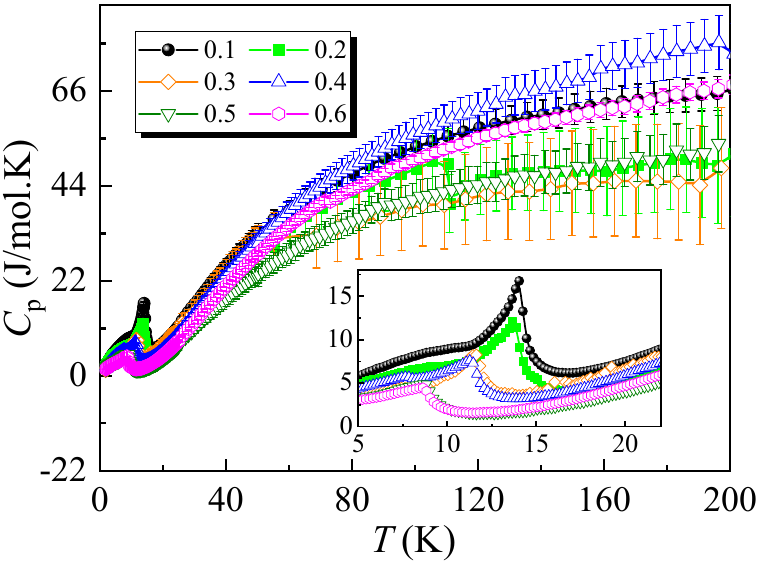}
\caption{The temperature-dependent variations in specific heat for different doping levels in Eu$_{1-x}$Sr$_x$AgP $(x = 0.0-0.6)$. The inset illustrates the specific shape and changes of the $\lambda$-shaped peak in the temperature range of 5--22 K.
}
\label{HC}
\end{figure}

The heat capacity results are depicted in Fig.~\ref{HC} with varying doping gradients. Pronounced $\lambda$-shaped transitions are observed around FM transition temperatures, signifying the occurrence of a magnetic phase transition. As the doping concentration increases, indicating a higher proportion of non-magnetic Sr$^{2+}$ ions, the initial FM order is suppressed, leading to a gradual decrease in the transition temperature and a diminishment of the intensity and significance of the $\lambda$-type peak simultaneously. The shift in this transition temperature corresponds with the alterations seen in the FM transition temperatures in our magnetization curves. While the flattening trend of the $\lambda$-shape peak indicates a gradual reduction in the contribution of magnons to the heat capacity, the predominant contributions are from phonons and electrons.\cite{tari2003specific, PhysRevB.68.174429} We conducted additional analysis of the specific heat data using a Debye-Einstein model, as depicted in Fig. S4.

\subsection{Resistivity study}
\begin{figure*}[t]
\centering
\includegraphics[width = 0.88\textwidth] {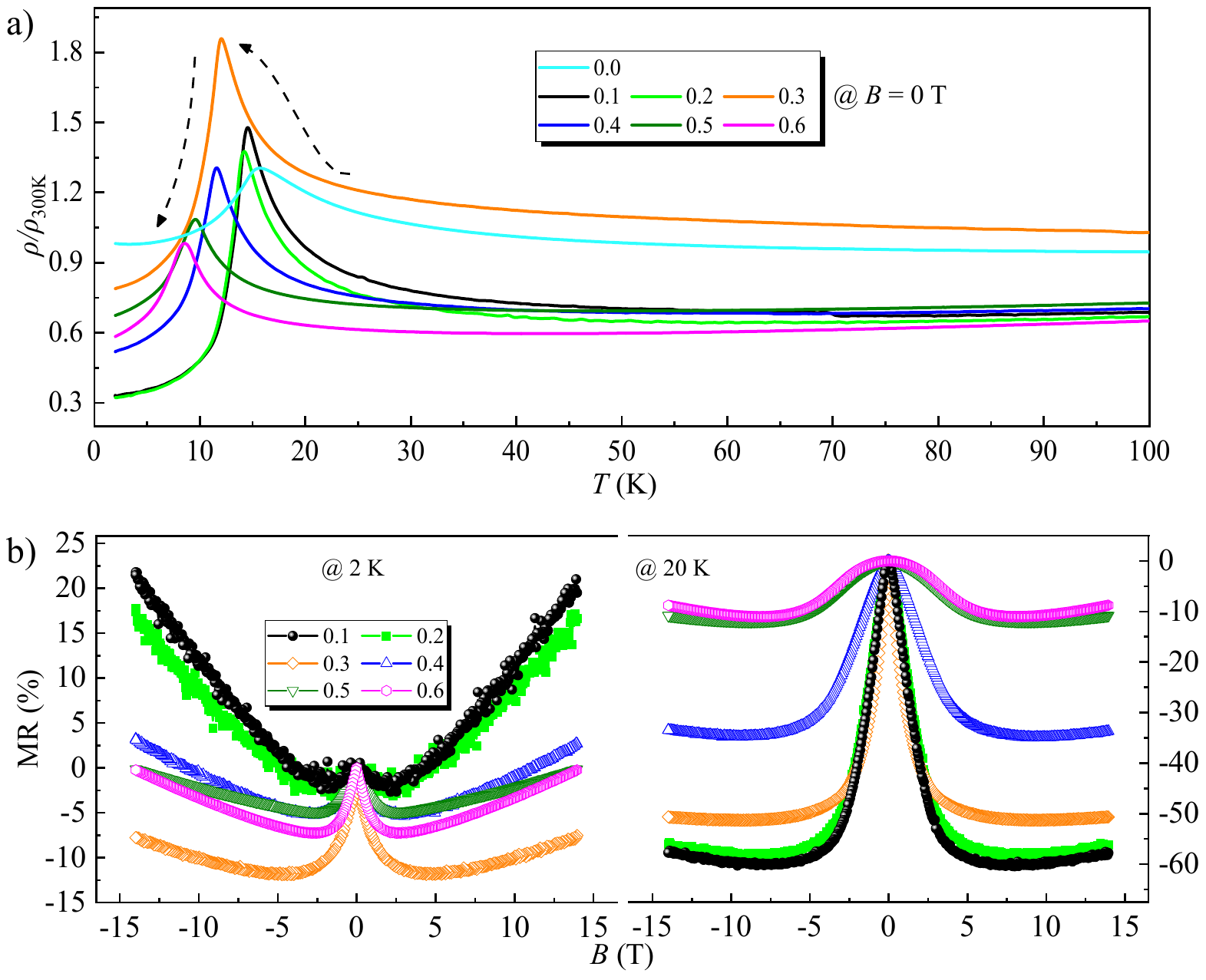}
\caption{Resistivity analysis and associated MR effect.
(a) The temperature-dependent resistivity at zero magnetic field, with the resistivity data normalized to $\rho_{\textrm{300K}}$, for Eu$_{1-x}$Sr$_x$AgP $(x = 0.0-0.6)$. Arrows indicate the shift trend of the resistivity maxima.
(b) Field dependence of the calculated magnetoresistance (MR\%) at 2 K and 20 K for Eu$_{1-x}$Sr$_x$AgP $(x = 0.0-0.6)$.
}
\label{Res}
\end{figure*}

As shown in Fig.~\ref{Res}a, we corrected the resistivity data by normalizing to $\rho_{\textrm{300 K}}$ to avoid the strong dependence of measured absolute resistance values on electrode quality. This correction was implemented to facilitate a more accurate comparison, emphasizing the effects of doping and temperature. The plotted data clearly reveals that, as the temperature decreases, Eu$_{1-x}$Sr$_x$AgP at various doping concentrations exhibits a more pronounced peak compared to the parent EuAgP compound. This peak signifies the magnetic ordering process and the development of the semiconductor state, i.e., spin-disorder scattering.\cite{PhysRev.168.531} As magnetic order is established, the carrier concentration gradually increases, indicating that Sr doping did not disrupt the FM semiconducting ground state of the samples. Simultaneously, the shifting of the peak position with the doping gradient aligns with our magnetic and specific heat measurements, confirming that Sr doping reduces the FM transition temperature. However, under zero magnetic field, Sr doping significantly enhances the temperature dependence of charge transport, especially at the doping concentration of 0.3. This may be related to the layered stacking of EuAgP. In each hexagonal unit cell of EuAgP, if precisely one-third of Eu$^{2+}$ is replaced by Sr$^{2+}$, these Sr$^{2+}$ ions will occupy specific positions within the cell, such as body-diagonal locations. This significantly impacts the scattering of phonons to electrons.\cite{coey2010magnetism, PhysRevB.106.045107}

The magnetoresistance was calculated at 2 and 20 K and presented as a function of the magnetic field applied perpendicular to the direction of electric current (Fig.~\ref{Res}b), using the following equation:
\begin{eqnarray}
\textrm{MR}(B) = \frac{\rho(B) - \rho(B = 0)}{\rho(B = 0)} \times 100\%.
\label{MR}
\end{eqnarray}
At 2 K, samples with $x$ = 0.1 and 0.2 exhibited unexpectedly large positive magnetoresistance ($\sim$ 20\%, at 14 T), which showed an approximately quadratic increase with the magnetic field. This phenomenon may stem from the applied magnetic field inducing a reduction in the average hopping length, attributed to a wave function contraction effect. Initially spherical wave functions assume an olive or cigar shape when a strong magnetic field is present, leading to a significant decrease in the interference of wave functions between localized states. As a consequence, the resistance of the sample escalates with the increasing magnetic field, resulting in positive magnetoresistance.\cite{MAJUMDER2021167941} For samples with $x$ = 0.3 to 0.6, the weak or incomplete establishment of magnetic order leads to the lagging of the aforementioned effect in competition with spin disorder or the decrease in carrier concentration induced by the magnetic field.\cite{PhysRev.168.531} Additionally, all samples displayed a crossover from negative to positive magnetoresistance, illustrating the competitive nature of the above-mentioned effects.

At 20 K, the temperature at which the FM transition is imminent, typically the magnetoresistance effect is most significant. Samples with low doping levels, $x$ = 0.1 and 0.2, exhibit substantial negative magnetoresistance ($\sim$ -60\%, at 14 T), almost twice that of the parent sample. It can be observed that low-level Sr doping effectively enhances the quality of the samples, particularly in the response of magnetic behavior. Additionally, we can finely tune the magnitude and even the sign of magnetoresistance by controlling the concentration of Sr doping.

\subsection{Phase diagram}
\begin{figure}[t]
\centering
\includegraphics[width = 0.48\textwidth] {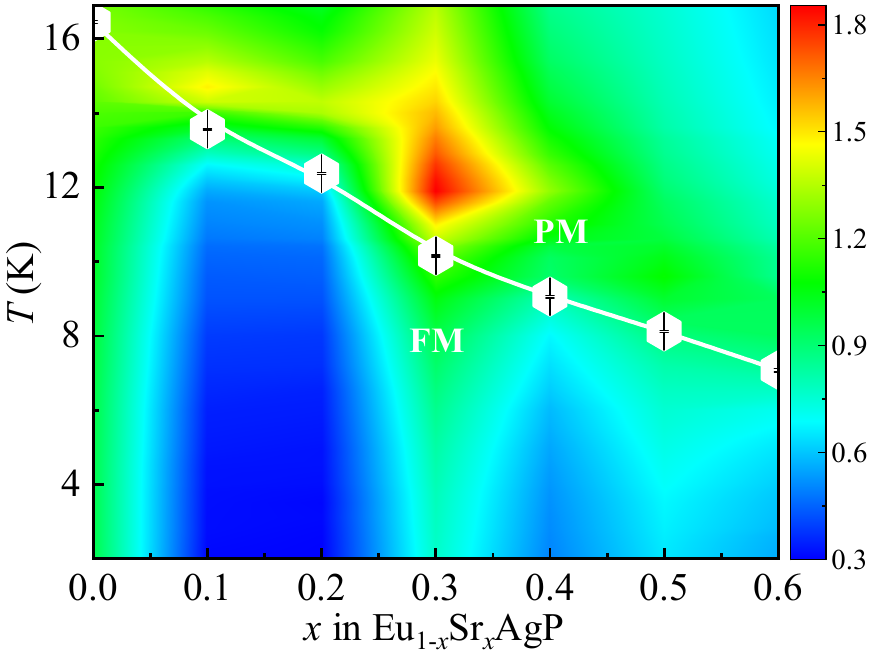}
\caption{Phase diagram of the magnetic phase transition from PM to FM in Eu$_{1-x}$Sr$_x$AgP $(x = 0.0-0.6)$ upon cooling, extracted from the temperature-dependent resistivity measurements. The hexagon symbols denote the FM transition temperatures at various doping levels. The color intensity represents the resistivity ratio of $\rho_{\textrm{T}}$/$\rho_{\textrm{300K}}$.}
\label{PD}
\end{figure}

Through the temperature-dependent $\rho/\rho_{\textrm{300 K}}$ and the correlation with varing doping levels, we constructed a phase diagram near the FM transition temperature and extracted the FM transition temperature ($T_{\textrm{C}}$) using CW fitting, as shown in Fig.~\ref{PD}. The transition temperature decreases with an increase in doping concentration, from the parent EuAgP sample ($T_{\textrm{C}}$ $\approx$ 16.45 K) to the Sr-doped Eu${_\textrm{0.4}}$Sr${_\textrm{0.6}}$AgP ($T_{\textrm{C}}$ $\approx$ 7.08 K), reflecting the inhibitory effect of Sr doping on magnetic exchange interactions. Notably, along the PM-FM transition temperature curve, the response of resistivity to temperature differs significantly on either side of the curve. In the PM region before the onset of magnetic order, the samples around $x$ = 0.3 exhibit a pronounced positive resistivity enhancement. In the FM region after the establishment of magnetic order, a general decrease in resistivity in the negative direction is observed in the phase diagram. This decrease is attributed to the reduction in phonon and spin fluctuations. However, these negative enhancements remain strongly correlated with the doping gradient. Such a phase diagram sheds light on precise control of resistivity in the EuAgP semiconductor.

\section{Conclusion}

In summary, the study of Sr-doped Eu$_{1-x}$Sr$_x$AgP $(x = 0.0-0.6)$ has provided valuable insights into the intricate interplay between structural modifications and diverse physical properties. Through systematic analyses, we observed the successful incorporation of Sr into the hexagonal crystal lattice of EuAgP, leading to expanded lattice constants. Magnetic studies revealed a suppression of the FM transition temperature with increasing Sr doping, highlighting the tunability of magnetic properties. The introduction of Sr doping with $x =$ 0.1 resulted in an effective PM moment of 7.83(1) ${\mu}_\textrm{B}$, markedly enhancing the magnetic robustness. Further validation of the impact of doping on magnetism was provided by specific heat measurements. Temperature-dependent resistivity exhibited distinct features, emphasizing the role of Sr doping in regulating charge transport. Magnetoresistance measurements showcased unexpected behaviors, demonstrating the tunability of magnetoresistance via Sr doping. Remarkably, the Sr doping resulted in both positive magnetoresistance, reaching up to 20\%, and negative magnetoresistance, reaching approximately -60\%. The tunability of magnetic and resistive properties through Sr doping highlights the potential for tailoring the properties of EuAgP. The constructed phase diagram provided a comprehensive view of the inhibitory effect of Sr doping on magnetic exchange interactions. Overall, the observed novel phenomena, including positive and negative magnetoresistance, offer opportunities for innovative applications in advanced technologies such as magnetic memory devices and magnetic sensor/actuator applications.

\section*{Author statement}

The manuscript “A new ferromagnetic semiconductor system of Eu$_{1-x}$Sr$_x$AgP $(x = 0.0-0.6)$ compounds: Crystallographic, magnetic, and magneto-resistive properties”, submitted for publication in Materials Today Communications, has not been published, and it is not being considered for publication elsewhere than Materials Today
Communications. 

\section*{CRediT authorship contribution statement}

\textbf{Qian Zhao:} Conceptualization, data curation, formal analysis, investigation, visualization, writing-original draft. \textbf{Kaitong Sun:} Formal analysis, investigation, methodology, visualization. \textbf{Junchao Xia:} Formal analysis, investigation, methodology, visualization. \textbf{Haifeng Li:} Conceptualization, funding acquisition, methodology, project administration, supervision, visualization, writing-review \& editing.

\section*{Declaration of competing interest}

The authors declare that they have no known competing financial interests or personal relationships that could have appeared to influence the work reported in this paper.

\section*{Data availability}

Data will be made available on request.

\section*{Acknowledgements}

This work was supported by the Science and Technology Development Fund, Macao SAR (File Nos. 0090{/}2021{/}A2 and 0049{/}2021{/}AGJ) and the Guangdong{-}Hong Kong{-}Macao Joint Laboratory for Neutron Scattering Science and Technology (Grant No. 2019B121205003).

\section*{Appendix A. Supporting information}
 
Supplementary data associated with this article can be found in the online version at . 

\bibliographystyle{elsarticle-num-names}
\bibliography{EuSrAgP}


\clearpage

\clearpage

\begin{table}[!t]
\small
\centering
\caption{Refined structural parameters of Eu$_{1-x}$Sr$_x$AgP from room-temperature XRPD, including lattice constants, unit-cell volume $V$, effective PM moment $\mu_{\textrm{eff}}$ from Curie-Weiss law fitting, and the goodness of refinements. The numbers in parenthesis represent the estimated standard deviations of the last significant digit.}
\label{ref-para}
\setlength{\tabcolsep}{5.8mm}{}
\renewcommand{\arraystretch}{1.1}
\begin{tabular} {llllllll}
\hline
\hline
\multicolumn{8}{c}{Eu$_{1-x}$Sr$_x$AgP $(x = 0.0-0.6)$}                                                            \\ [1pt]
\multicolumn{8}{c}{Hexagonal, space group $P6_3/mmc$ (No. 194)}                                                    \\ [1pt]
\hline
$x$           & $a$ (= $b$)           & $c$              & $V$                   & $\mu_{\textrm{eff}}$   & $R_\textrm{p}$   & $R_\textrm{wp}$       & $R_\textrm{exp}$         \\
              & (\AA)                 & (\AA)            & ({\AA}$^3$)           & (${\mu}_\textrm{B}$)   &                  &                       &                          \\
0             & 4.3937(1)             & 8.0587(2)        & 134.733(5)            & 6.21(1)                & 5.04             & 6.80                  & 3.08                     \\
0.1           & 4.3962(1)             & 8.0746(2)        & 135.139(5)            & 7.83(1)                & 6.68             & 9.93                  & 2.62                     \\
0.2           & 4.3970(1)             & 8.0922(2)        & 135.495(5)            & 6.84(1)                & 4.73             & 7.44                  & 2.51                     \\
0.3           & 4.3987(1)             & 8.1114(3)        & 135.919(7)            & 7.07(1)                & 5.68             & 9.46                  & 2.42                     \\
0.4           & 4.3999(1)             & 8.1291(2)        & 136.293(6)            & 7.39(1)                & 4.92             & 8.24                  & 2.31                     \\
0.5           & 4.4013(1)             & 8.1455(2)        & 136.656(4)            & 7.41(1)                & 5.38             & 8.57                  & 2.72                     \\
0.6           & 4.4015(1)             & 8.1559(2)        & 136.840(5)            & 7.14(1)                & 6.51             & 10.7                  & 2.69                     \\
\hline
\hline
\end{tabular}
\end{table}

\clearpage

\clearpage

\clearpage

\clearpage

\end{document}